\documentclass{IEEEcsmag}

\usepackage[colorlinks,urlcolor=blue,linkcolor=blue,citecolor=blue]{hyperref}
\usepackage{upmath}

\usepackage[utf8]{inputenc}
\usepackage{fontawesome}
\usepackage{colortbl}
\usepackage{calc}
\usepackage{graphicx}
\usepackage{booktabs}
\usepackage{tabularx}
\usepackage{microtype}
\usepackage{multirow}
\usepackage{subcaption}
\usepackage{tabularray}

\newcommand\figurewidth{0.24\textwidth}

\def\plaintitle{Mastery Learning Improves Performance on Complex Tasks on PCP Literacy Test}

\begin{document}

\title{\plaintitle{}}

\author{C. Srinivas$^1$, E.E. Firat$^2$, R. S. Laramee$^3$, A. P. Joshi$^1$}
\affil{$^1$University of San Francisco, $^2$Cukurova University, $^3$University of Nottingham}

\begin{abstract}
Developing literacy with unfamiliar data visualization techniques such as Parallel Coordinate Plots (PCPs) can be a significant challenge for students. We adopted the Revised Bloom's taxonomy to instruct students on Parallel Coordinate Plots (PCPs) using Mastery Learning in the classroom. To evaluate Mastery Learning's impact, we conducted an intervention in a Data Visualization course to teach students about PCPs using the Revised Bloom's taxonomy with and without Mastery Learning. Based on our intervention, we found that while students in both groups performed similarly on the first two (Remember, Understand) modules, the students in the Mastery Learning group performed better on modules that required more advanced thinking (Analyze, Evaluate) and demonstrated a better comprehension of PCPs. We provide all the materials developed including the six-module Bloom's Taxonomy PCP literacy (BTPL) test for full reproducibility on our website at \protect\url{https://vis-graphics.github.io/PCP-Literacy-Test/.} 
\end{abstract}

\maketitle{}

\begin{figure*}[t]
  \centering 
   \includegraphics[width=\textwidth]{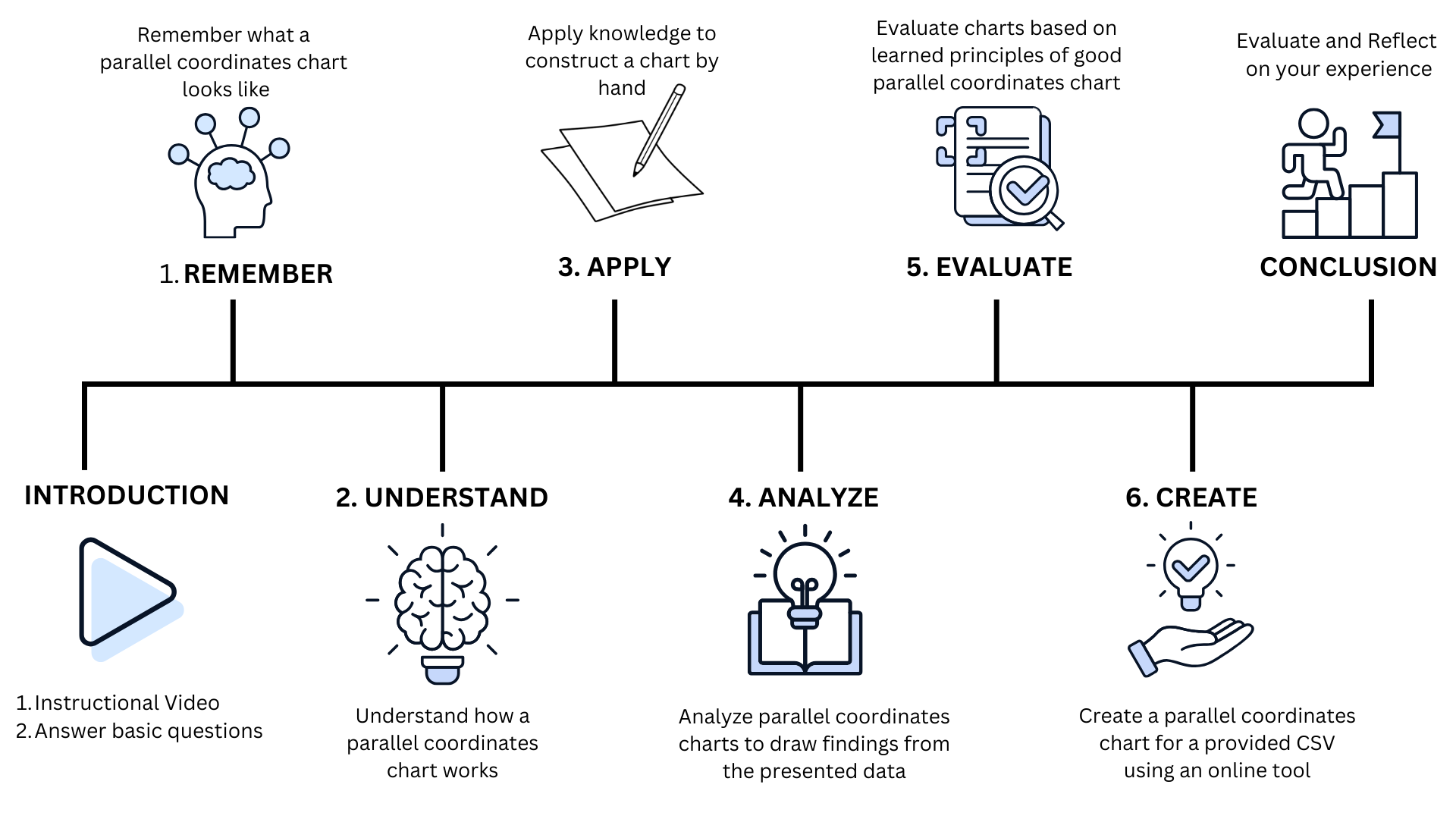}
\caption{This figure shows the pathway of intervention that includes Revised Bloom's taxonomy-based modules to teach students PCPs.}
\label{fig:modules}
\end{figure*}

\section{Introduction}

As a society, we increasingly rely on visual representations of data to make informed decisions about health, home energy usage, finances and so on. The ability to effectively read, interpret, and critically evaluate graphical representations of data has become extremely crucial. While visualization literacy has traditionally been defined as the ability to read and interpret charts~\cite{borner2019vis}, language literacy is defined as the ability to confidently read, speak, \textit{and} write in a language~\cite{montoya2018defining}.

In this paper, we present the results of our intervention with students in a Data
Visualization course that are taught to read, interpret, analyze, and create Parallel Coordinate Plots (PCPs). 
The teaching material was created by following the principles of the Revised Bloom's taxonomy by Kratwohl~\cite{Krathwohl01112002}. The Revised Bloom's taxonomy contains six stages of increasingly complex cognitive processes: \textit{Remember, Understand, Apply, Analyze, Evaluate,} and \textit{Create}.

\textit{Mastery Learning (ML)} is a pedagogical strategy that requires students to demonstrate competency on a particular learning module, before continuing onto the next increasingly difficult module~\cite{block1976mastery}. 
In this paper, we present the results of evaluating the impact of ML on student performance using the Revised Bloom's taxonomy-based modules that focus on the topic of Parallel Coordinate Plots (PCPs). The contributions of our paper are as follows:

\begin{enumerate}
\item Mastery Learning (ML) modules based on Bloom's taxonomy to develop PCP literacy,
\item Assessment of the impact of ML when teaching students about PCPs using the modules,
\item An Open PCP Literacy Test (\textbf{Bl}oom's taxonomy based \textbf{P}CP \textbf{L}iteracy Test - BTPL) \footnote{\url{https://vis-graphics.github.io/PCP-Literacy-Test/}} with the content and the assessments per module provided for reproducibility and potential reuse by instructors in their own data visualization courses. 
\end{enumerate}

\section{Related Work}


Visualization literacy is defined by B\"{o}rner et al. as, ``the ability to make meaning from and interpret patterns, trends, and correlations in visual representations of data''~\cite{borner2016investigating,borner2019vis}, while Lee et al.~\cite{lee2016vlat} referred to it as ``the ability and skill to read and interpret visually represented data and to extract information from data visualizations.'' Firat et al.~\cite{firat2022interactive} present a literature review that discussed various studies from the data visualization and related communities that investigate, evaluate, and test visualization literacy skills.
Assessments for overall visualization literacy have been developed such as VLAT~\cite{lee2016vlat}, mini-VLAT~\cite{pandey2023mini}, CALVI~\cite{ge2023calvi}. We present a PCP Literacy Test (BTPL) that can be used to teach and assess the ability of learners on various aspects of PCPs. 



\begin{figure}[!t]
    \centering
    \includegraphics[width=.75\columnwidth]{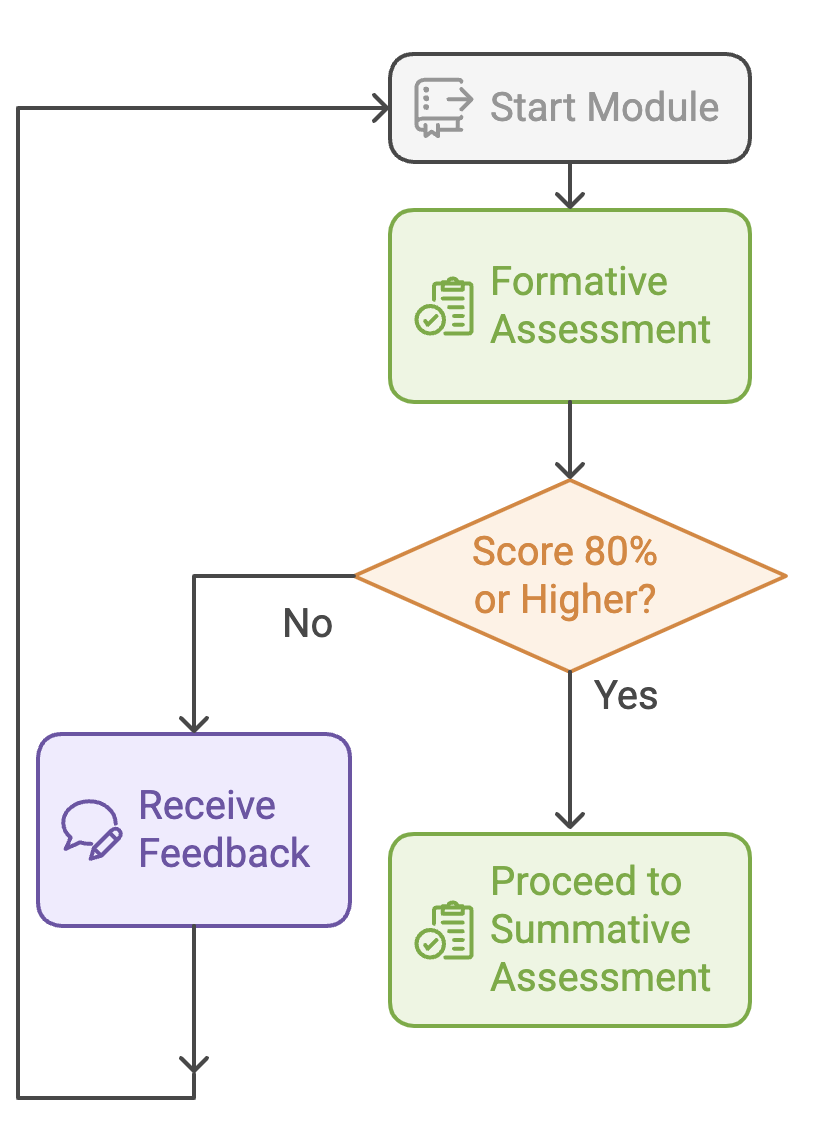}
    \caption{This figure demonstrates the procedure that the students in the ML group follow. Students in the Control group move to the Summative Assessment without any feedback.}
    \label{fig:repeat-criteria}
\end{figure}

Bloom's taxonomy was \textbf{revised} by Kratwohl~\cite{krathwohl2002revision} to include the learner's thinking \textit{processes} rather than behaviors. Burns et al.~\cite{burns2023we} developed a framework utilizing Bloom's taxonomy to assess different levels of understanding of charts. They found it to be superior to existing methods focused solely on quick and accurate comprehension. 
Adar and Lee applied Bloom's taxonomy to help designers identify learning objectives for better defining, assessing, and comparing communicative visualizations, but they did not specifically target increasing students' visualization literacy for a particular technique~\cite{adar2020communicative}. Our approach differs from previous work as we \textit{investigate the benefits of incorporating Mastery Learning (ML) into modules teaching Parallel coordinate plots} that are developed with Bloom's taxonomy.

Adaptive learning systems customize the learning journey according to each student's unique requirements and capabilities. 
Garner et al.~\cite{garner2019mastery} found studies, which recorded the usage of mastery learning in computer science courses and their positive results based on observations and comparative performance. 

Prior research in data visualization literacy has primarily concentrated on testing various educational methods to instruct individuals on specific techniques. 
Peng et al.~\cite{peng2022} provided Corrective Immediate Feedback (CIF) to students as they were learning new concepts about PCPs. They found that students who received immediate feedback outperformed those without feedback across all modules. In this work, we specifically explore the impact of \textit{ML}~\cite{block1976mastery} and the impact it has on the performance of students as they study PCPs. 

\section{Approach}

We developed and evaluated the impact of ML-based modules on student understanding of PCPs based on the material by Peng et al.~\cite{peng2022} that served as a starting point for the modules. Those modules are based on the Revised Bloom's taxonomy principles, but do not contain any of the ML implementation. According to ML principles~\cite{block1976mastery}, students see the results of the
assessment \textit{at the end of the assessment} and are required to retake the
assessment after receiving feedback and reviewing the material (text, videos, and so on). We incorporated these principles into the 
the Remember, Understand, Analyze, and Evaluate modules of the Revised Bloom's taxonomy such that students must retake the 
Formative Assessment (FA), based on their scores in the FA.

Figure~\ref{fig:modules} shows an overview of the modules and the pathway through the entire intervention. Students complete each module in order of increasing complexity, using the cognitive processes prescribed in the Revised Bloom's taxonomy. At every step, they learn more in-depth about PCPs and then complete the Formative Assessment (FA) and the Summative Assessment (SA) (See Figure~\ref{fig:repeat-criteria}).
Students in the ML group receive feedback on their answers for all the questions in that module after completing an attempt of the FA. If they score less than 80\% on the FA, they retake the FA after reviewing the instructional material before they continue to the SA. Students in the control group do not receive any feedback after the formative assessment and continue on to the SA regardless of their score on the FA for each module. The 80\% threshold was determined based on a literature review of ML~\cite{block1976mastery}.

\subsection{Mastery Learning Modules}

\begin{table*}
\begin{tblr}{
  colspec = {Q[l,co=0]X[l]X[c,h]Q[l,co=0]Q[l,co=0]},
  stretch = 0,
  rowsep = 5pt,
  hlines = {1pt},
  vlines = {1pt},
}
         \textbf{Cognitive Process} & \textbf{Description} &\textbf{Representative Figure} & \textbf{Formative} & \textbf{Summative} \\ \hline 
         Remember & \parbox{\dimexpr\linewidth-2\fboxsep-2\fboxrule\relax}{Recognize a parallel coordinates chart individually or from a line up of other charts } & \includegraphics[width=\figurewidth]{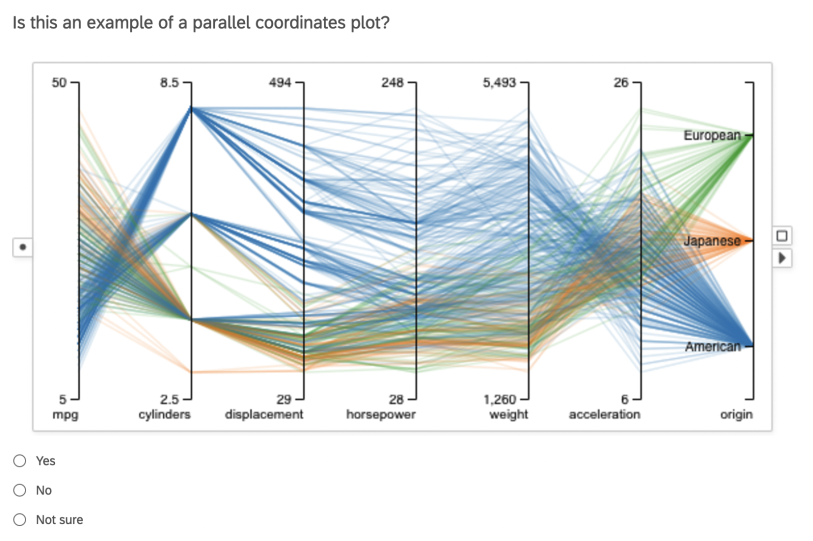} & 8 & 9 \\ 
         Understand & Understand the various characteristics of a PCP related to identifying axes, learning about correlations among variables of neighboring axes, and axis reordering. & \includegraphics[width=\figurewidth]{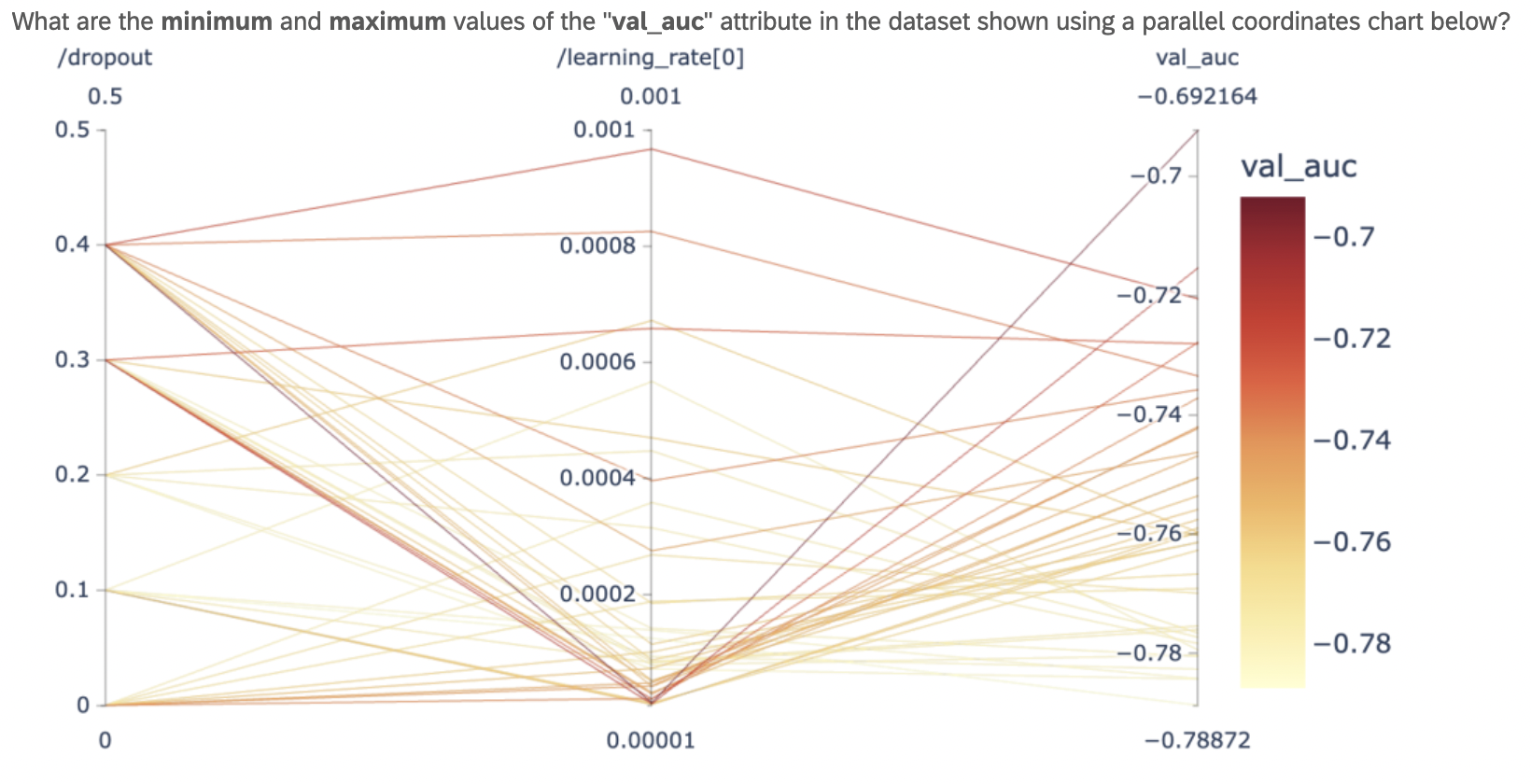} & 10 & 9 \\ 
         Apply & Draw a PCP on paper or using an electronic drawing tool. Students answer 5 questions using the PCP drawn by them & \includegraphics[width=\figurewidth]{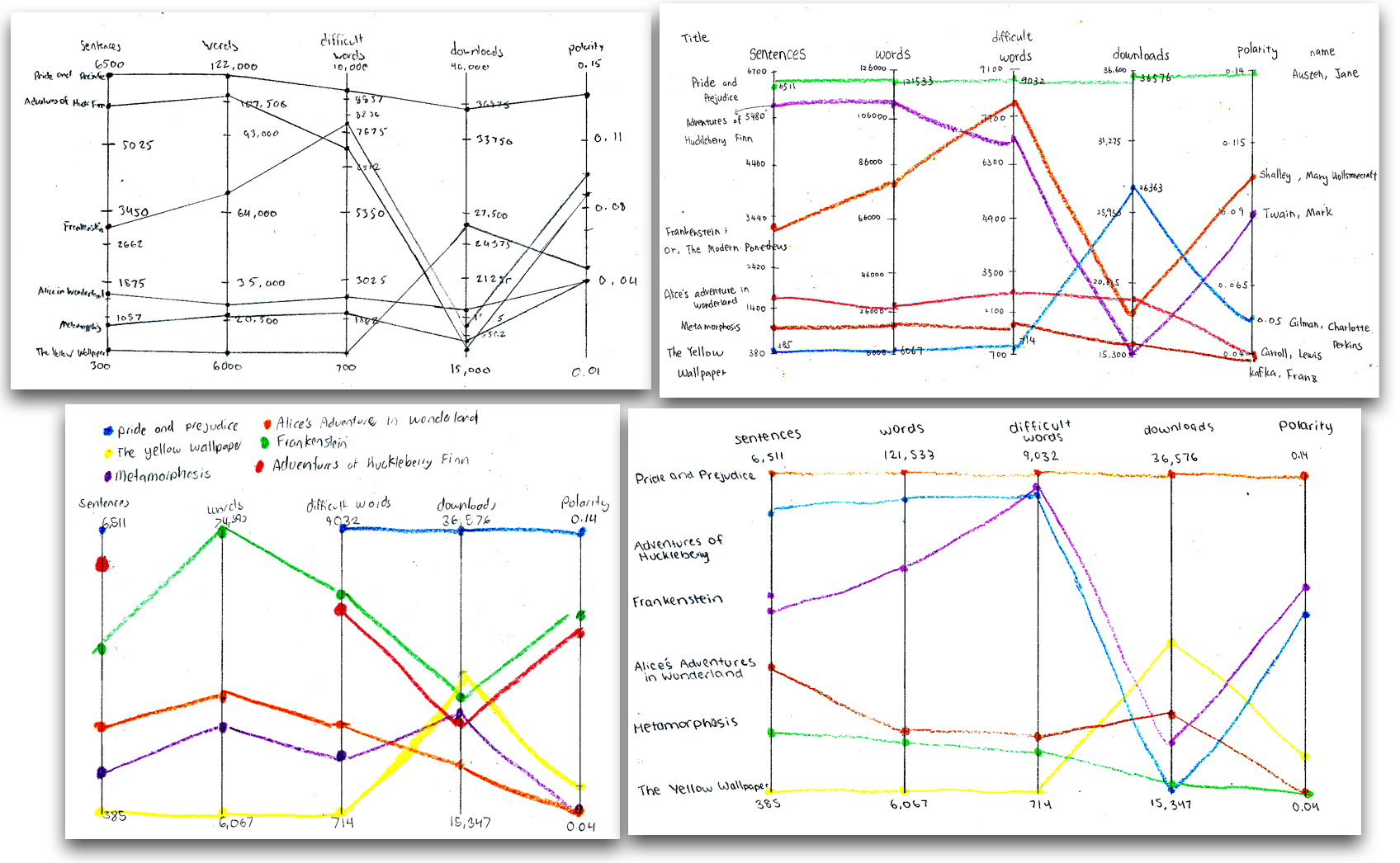} & -  & 5 \\
         Analyze & \textit{Make decisions} about a scenario using PCPs that requires \textit{finding patterns} in the data across multiple axes. Some questions require tracing polylines across multiple axes to answer queries. & \includegraphics[width=\figurewidth]{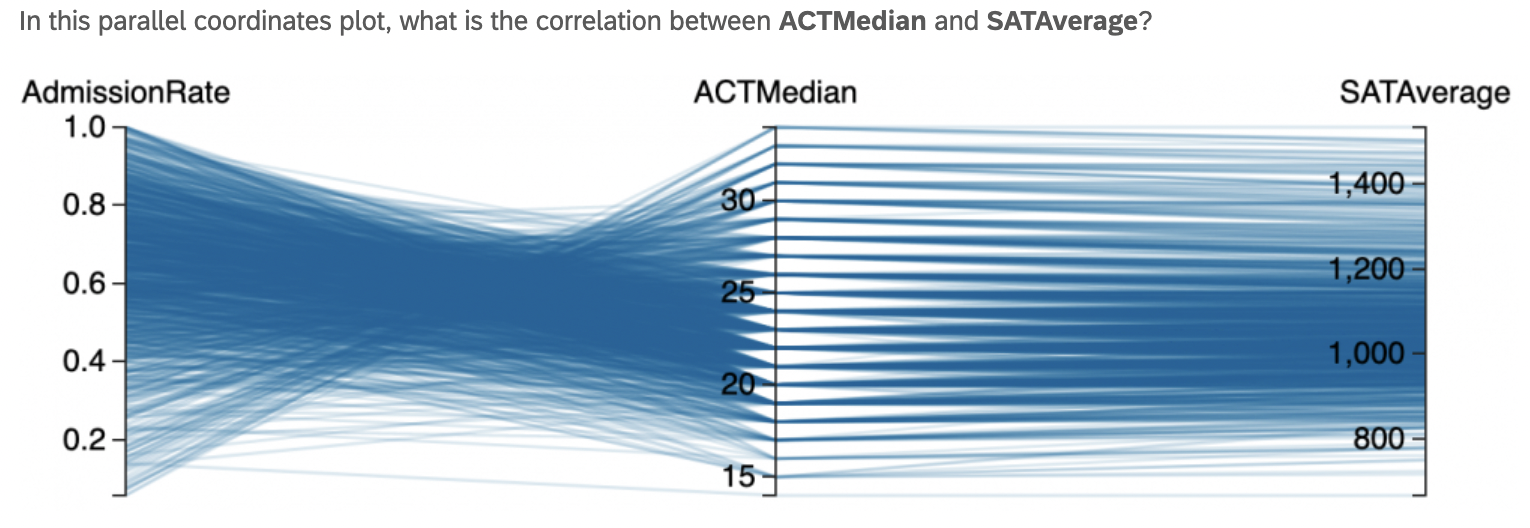} &  8 & 9 \\ 
         Evaluate & \textit{Critique} and \textit{Evaluate} the efficacy of the PCPs that contain flaws in them (missing axis labels, mismatched color legends, repeated axes, and so on.) & \includegraphics[width=\figurewidth]{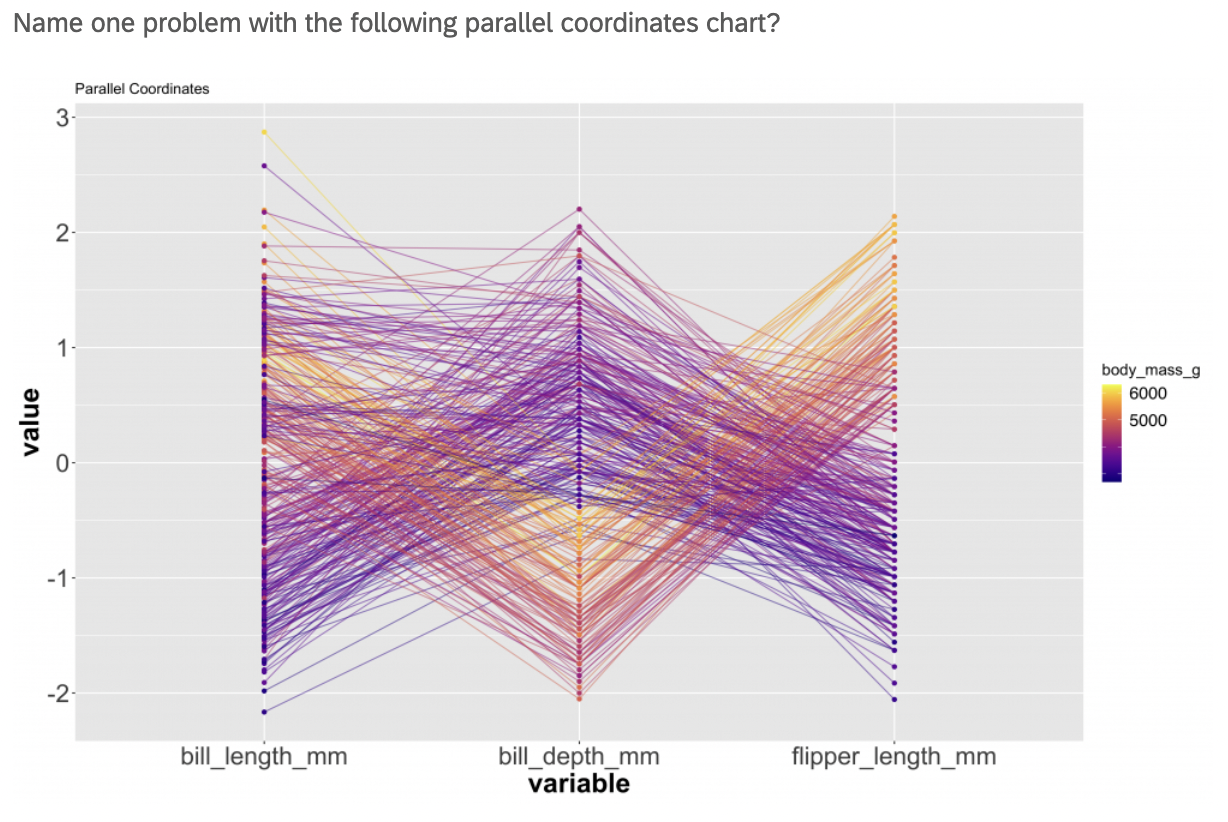} & 15 & 12\\ 
         Create & \textit{Create} a PCP using RawGraphs.io and an online tool using the d3.parcoords package. Students answer 5 questions that based on PCP generated by them. & \includegraphics[width=\figurewidth]{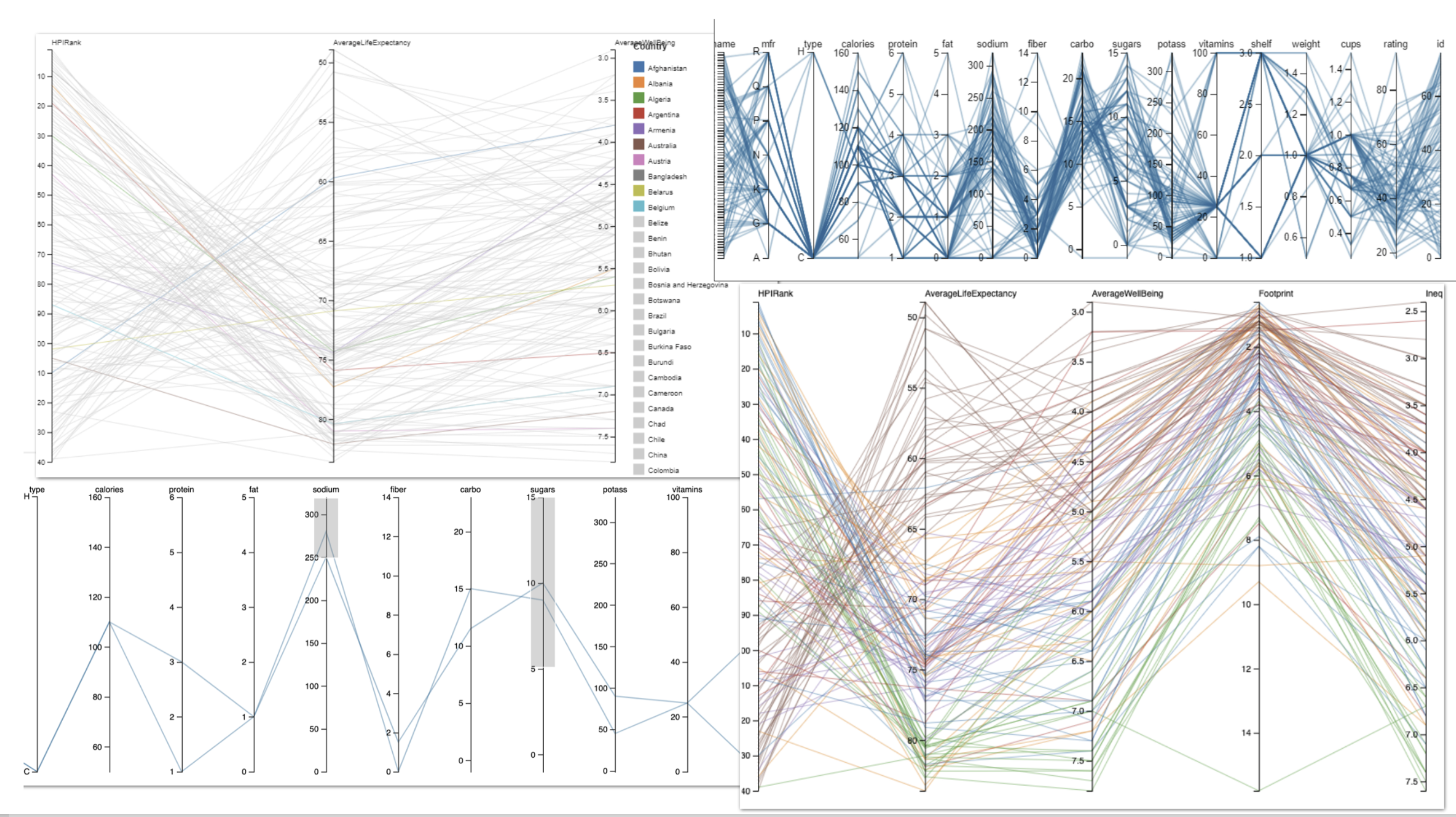} & - & 5 \\ 
\end{tblr}
    \caption{This table contains an overview of the Cognitive Processes from the Revised Bloom's taxonomy including the number of questions that comprised the FA and SA in each respective module. Each question asked at each stage of the cognitive process is available in the supplementary material at \url{https://github.com/vis-graphics/ml-pcp-literacy}}
    \label{tab:assessments}
\end{table*}


Each module follows the cognitive processes in the Revised Bloom's taxonomy in increasing complexity. Table~\ref{tab:assessments} shows an overview of the number of questions in the formative and summative assessment for each cognitive process. There are no FA questions for the Apply and Create modules due to the fact that students are drawing PCPs (by hand or using an online tool) and the learning occurrs in an applied setting in these cognitive processes. ML was incorporated into the other four processes (Remember, Understand, Analyze, and Evaluate) such that the students (in the ML group) saw their score on the FA after every attempt of the FA. The difficulty level for the \textit{Remember} and \textit{Understand} modules is \textit{much lower}, as all the students have to do is identify what a PCP looks like and recognize the various components of a PCP (axis labels, axis ordering, and so on). The \textit{Analyze} and \textit{Evaluate} modules require students to apply \textit{higher-order thinking skills} to complete the assessment. This was validated by the comparative performance of the students on the modules as well. These findings are discussed in detail in the Results Section.  

All the questions asked in each cognitive process are available in the supplementary material at \url{https://github.com/vis-graphics/ml-pcp-literacy} as well as on the PCP Literacy Test website (BTPL) \footnote{\url{https://vis-graphics.github.io/PCP-Literacy-Test/}}.

\subsection{Intervention Details}

To evaluate the impact of ML on students' understanding of PCPs, we conducted an intervention with undergraduate students in a data visualization course. Students were assigned either to the \textit{ML} group or the \textit{Control} group. There were a total of 55 undergraduate students across two different offerings of a Data Visualization course. Out of those, 54 students were in the 18-24 age group, 1 student was in the 25-44 age group with 14 females and 41 males. 27 students were in the ML group and 28 were in the Control group. The intervention was carried out in a single session lasting a little more than two hours on average. Students in both the groups then completed the SA after each module. We compare the students' performance on the SA to evaluate the impact of ML on their learning for the given module. 



\section{Results}
\label{sec:results}

For the analysis, we examined the accuracy of students on individual questions in each SA (Fig.~\ref{fig:results-composite}) as well as examined the overall scores received by the students in each SA (Figure~\ref{fig:modules-summary}). We discuss each module as well as specific questions where majority of the students in one group (or both groups) performed poorly. 


\subsection{Remember Module}

\begin{figure}
    \centering
    \includegraphics[width=1.15\columnwidth]{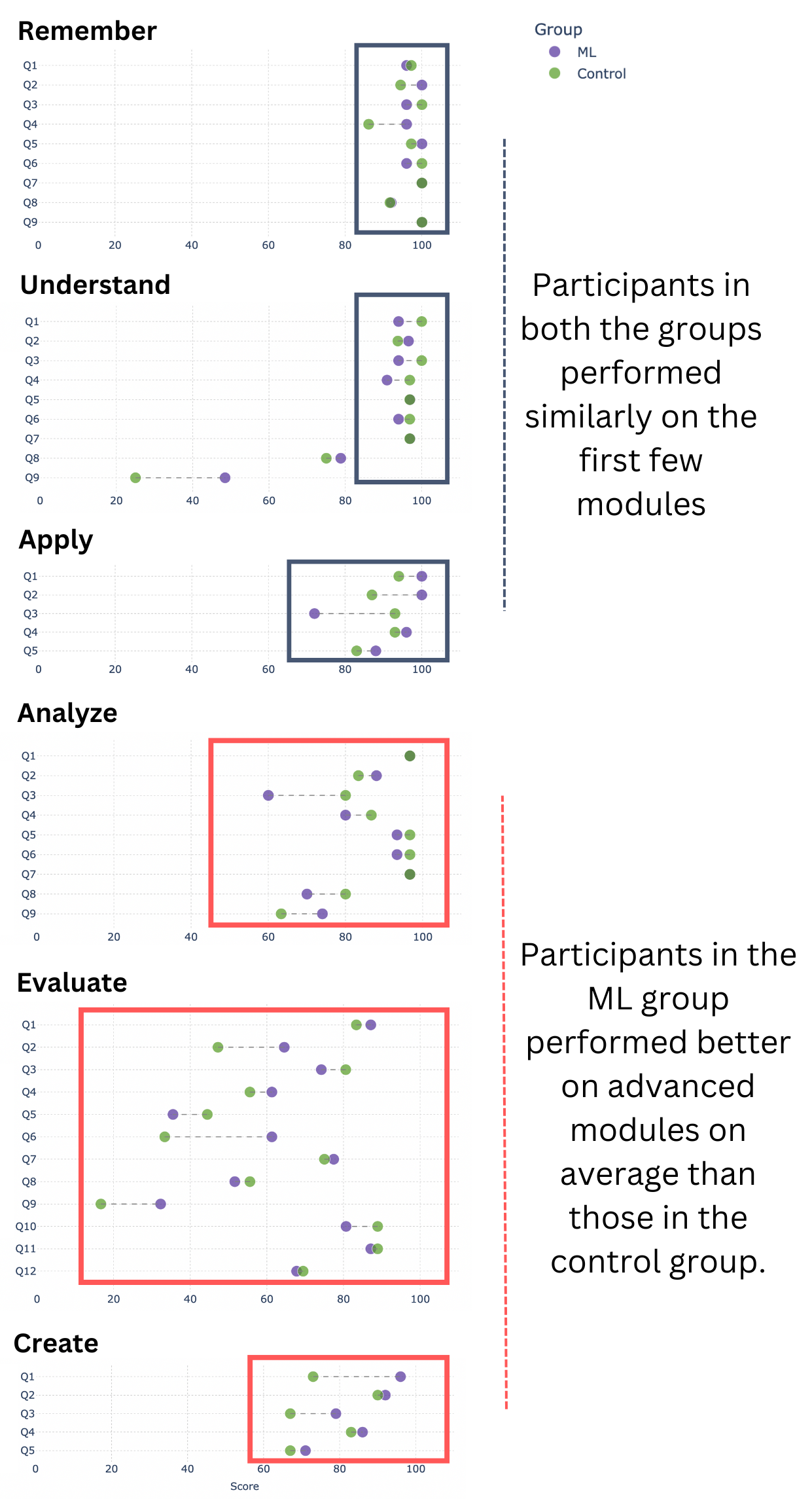}
    \caption{Students in both the groups performed equally well for the first three modules, but the students in the ML group performed better on the advanced modules.}
    \label{fig:results-composite}
\end{figure}

\begin{figure*}%
    \centering
    \subfloat[\centering Q4 from Remember SA]{{\includegraphics[width=0.49\textwidth]{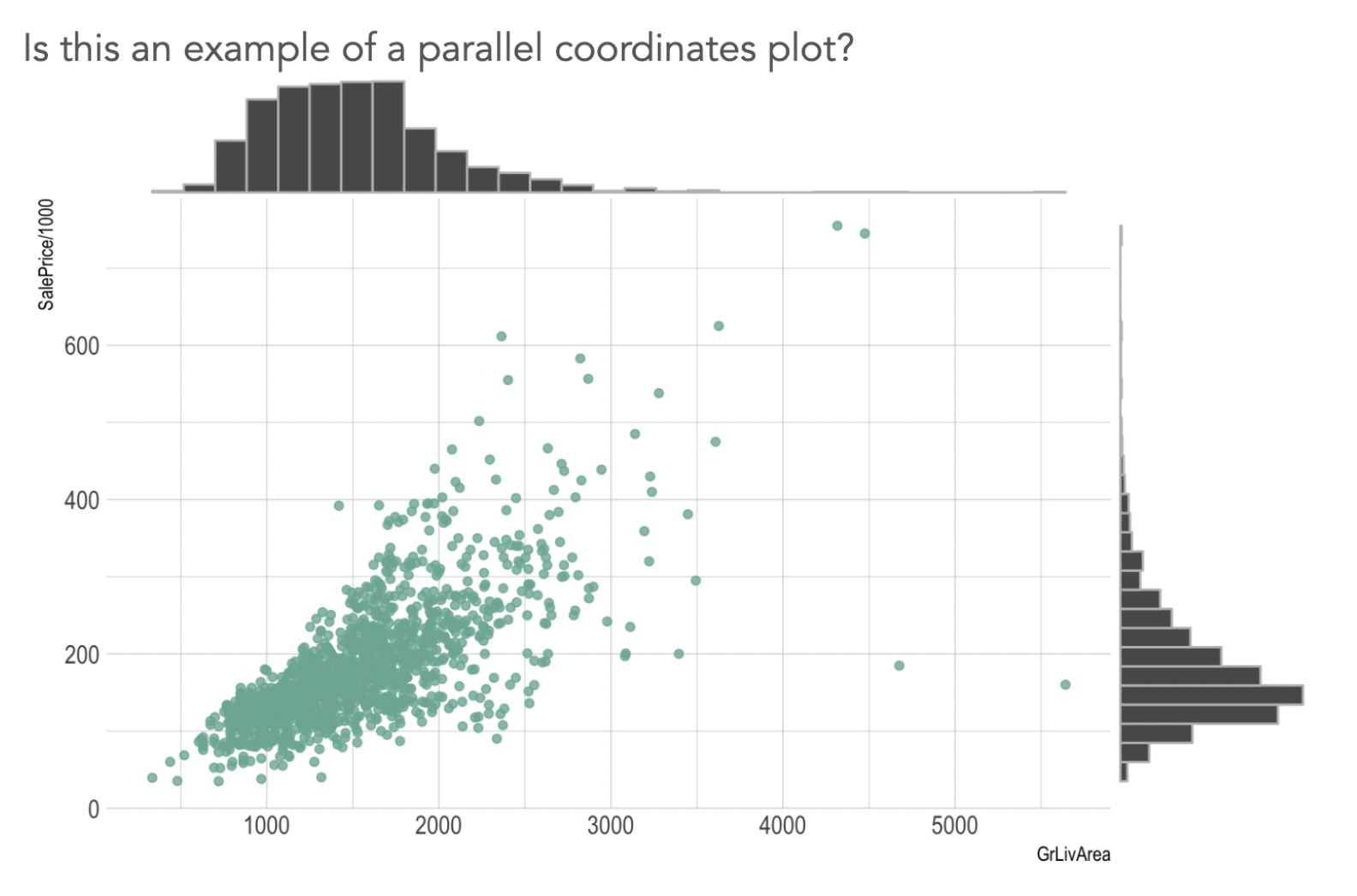} }}%
    \subfloat[\centering Q9 from Understand SA]{{\includegraphics[width=0.49\textwidth]{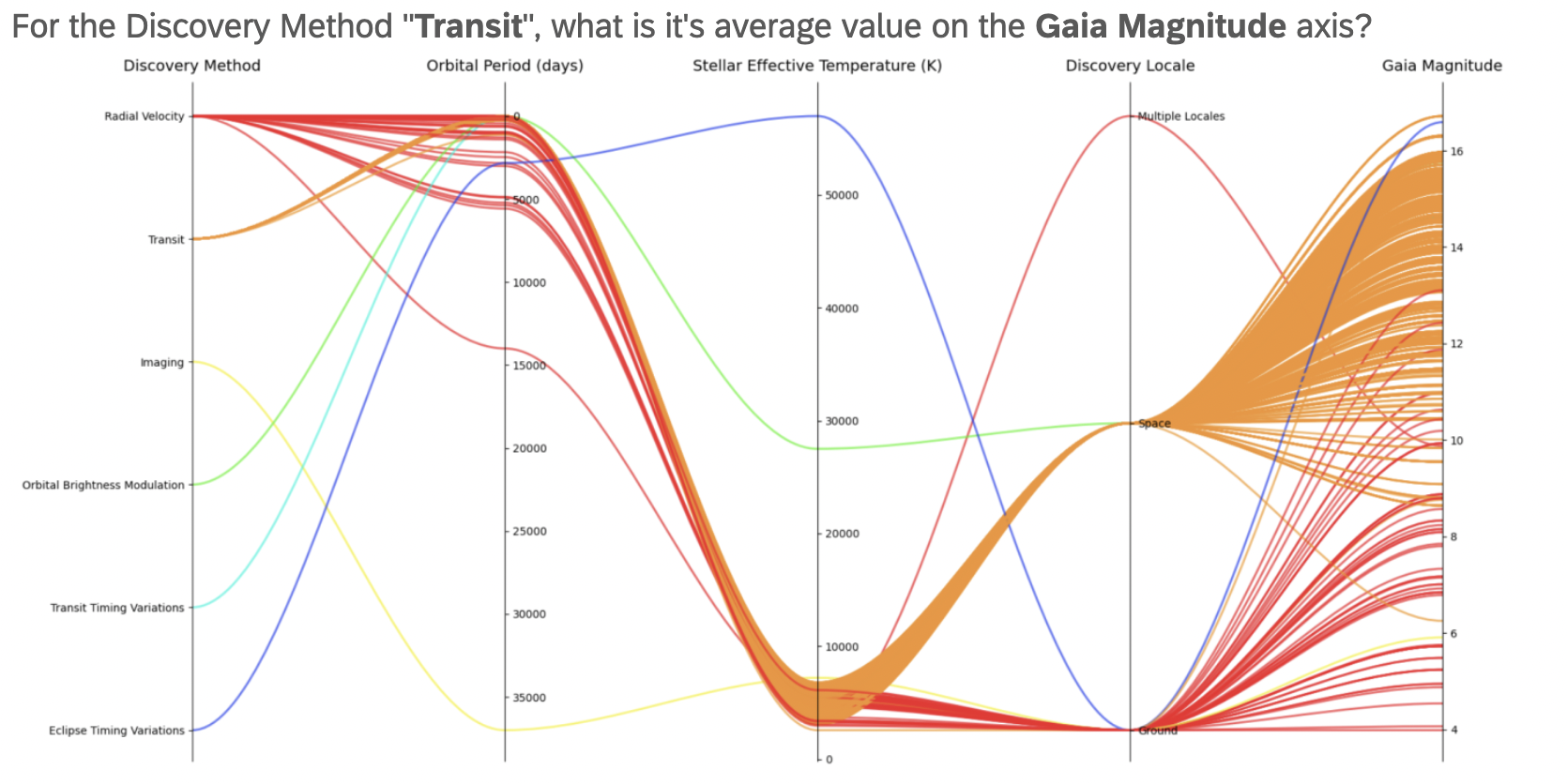} }}%
    \\
    \subfloat[\centering Q9 from Analyze SA]{{\includegraphics[width=0.49\textwidth]{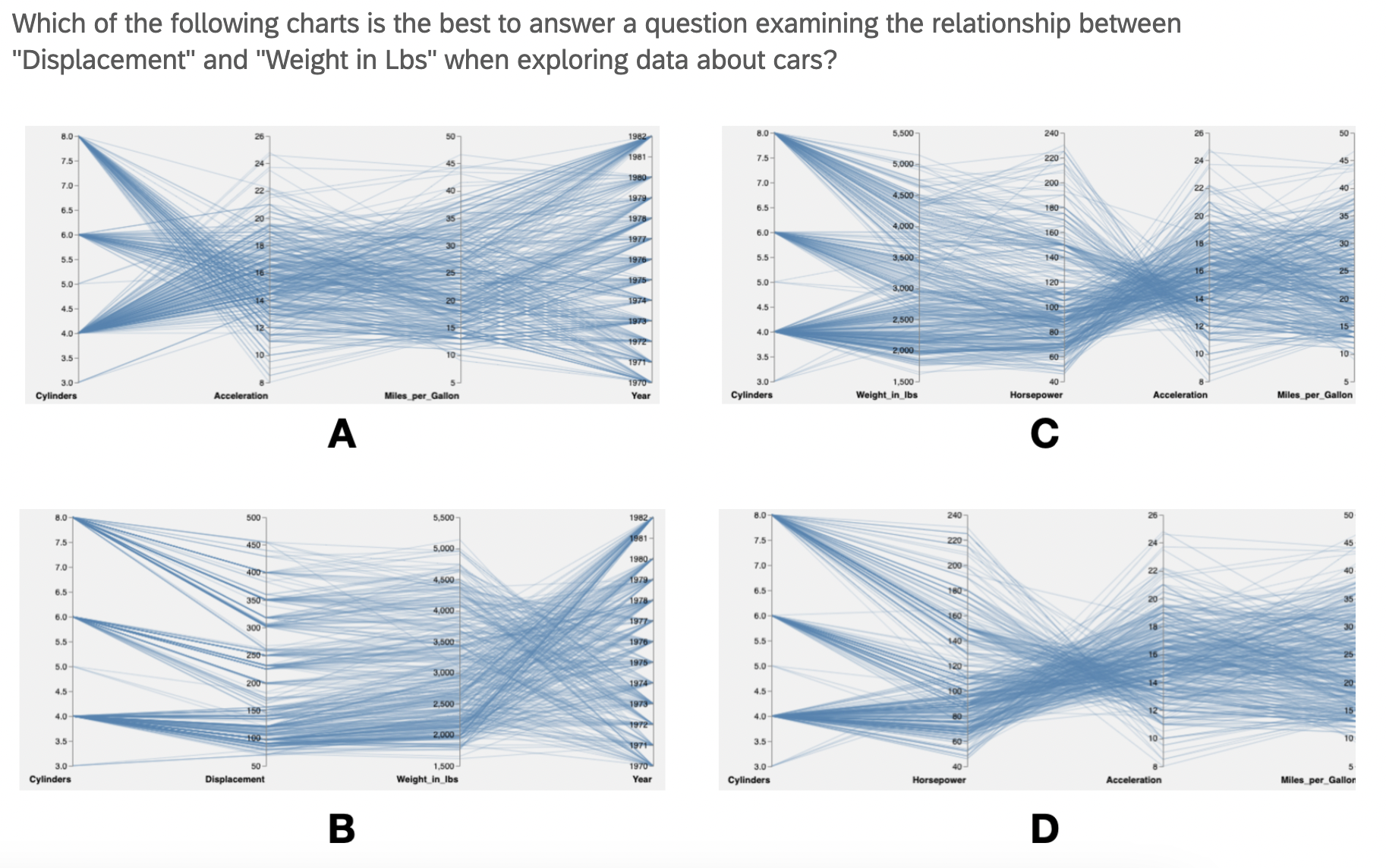} }}%
    \subfloat[\centering Q9 from Evaluate SA]{{\includegraphics[width=0.49\textwidth]{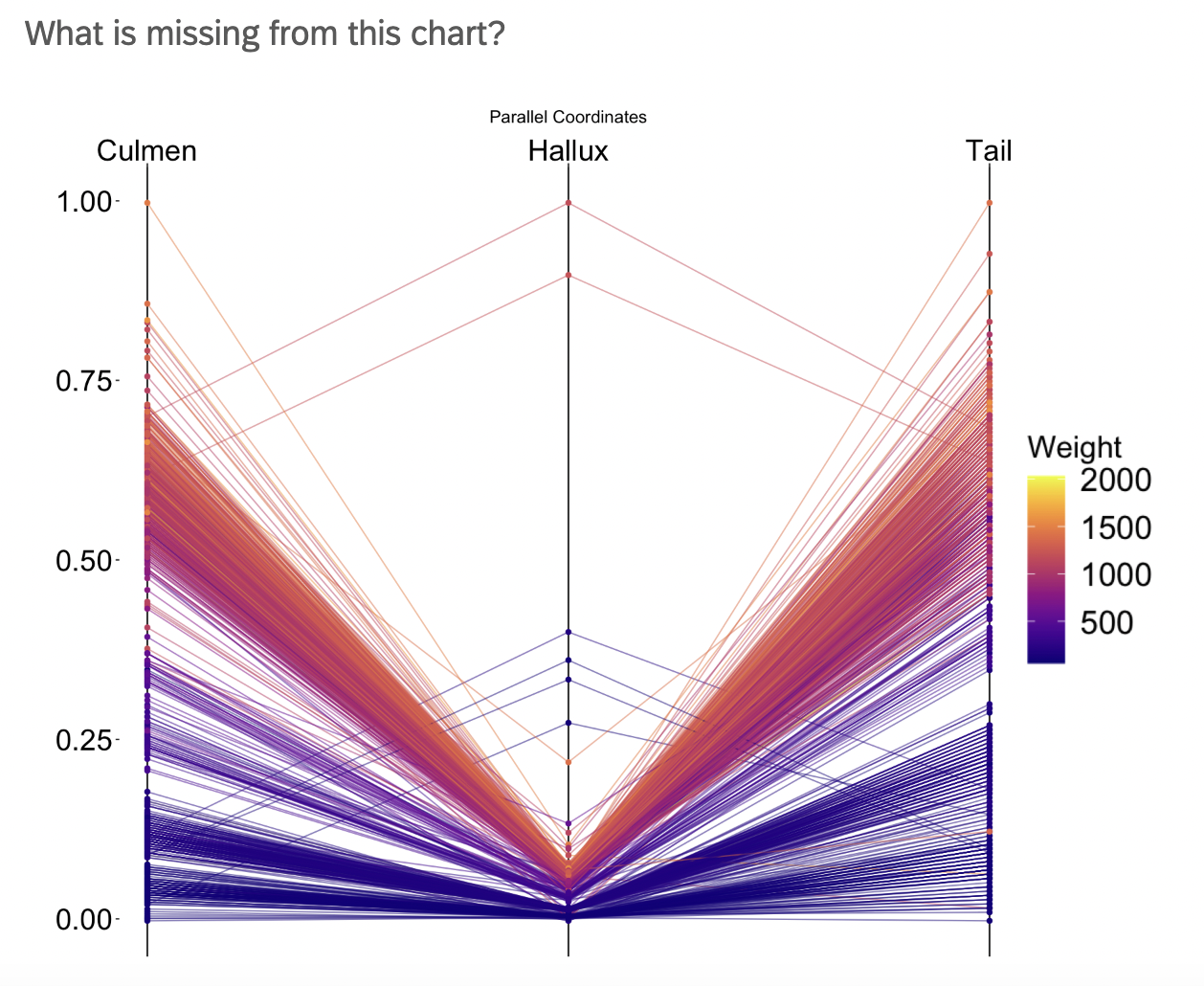} }}%
    \caption{Difficult Questions in each module - Top left figure shows Question 4 from the Remember SA, where 96\% in the ML group answered it correctly compared to 86\% in the Control group. The Top right figure shows Q9 from the Understand SA that students in both the groups answered incorrectly. Students may have confused this chart with a Sankey diagram~\cite{srinivas2024inductive}. Bottom left figure shows four PCPs that the students had to choose from to answer a question about two data variables in the PCP. Bottom right figure shows Q9 from the Evaluate SA that students in both the groups struggled to critique. The chart was missing axis labels and the color legend does not start at zero.}%
    \label{fig:all_questions}%
\end{figure*}

The SA of the Remember module consists of 9 questions that includes identifying whether a chart was a PCP or not and identifying a PCP out of a line-up of charts. The topmost chart in Figure~\ref{fig:results-composite} shows the performance of students on individual questions in both the groups. A large majority of the students (greater than 90\%) in both the groups (ML = purple, Control = green) answered the questions accurately. Question \textit{Q4} is one question where students in the control group performed worse than those in the ML group. Figure~\ref{fig:all_questions}(a) shows the exact chart that students saw. Students in the control group may have misinterpreted the scatterplot as a PCP~\cite{srinivas2024inductive}.


\subsection{Understand Module}

The SA of the Understand module consists of 10 questions. The second chart from the top in Figure~\ref{fig:results-composite} shows
the performance of the students in both the groups on the SA. For most questions, a large majority of the students in both the groups answered the questions well. For Q9, though, only 25\% of the students in the Control group answered it correctly, whereas 48\% of the students in the ML group answered the question correctly. Figure~\ref{fig:all_questions}(b) shows a screenshot of that question. We conjecture that students in both the groups may have misread the chart and could potentially have confused it for a similar visualization design such as a Sankey diagram or a line graph~\cite{srinivas2024inductive}.

\subsection{Apply Module}

For this module, students applied their knowledge by drawing a PCP using pen-and-paper or a digital tool. They then answered 5 questions using the PCP that they had drawn by hand. The third chart from the top in Figure~\ref{fig:results-composite} shows that the majority of the students are able to answer most of the questions accurately in both the groups. 

Additionally, we also graded each hand-drawn chart on a 10-point rubric (included in supplementary material) for their completeness, use of color, data labels, axis labels, and so on. The median score for the students in the ML group was higher (\textit{9/10}) as compared to the median score in the Control group (\textit{8/10}).

\subsection{Analyze Module}

The SA of the Analyze module contains 9 questions that students in both groups answered. The fourth chart in Figure~\ref{fig:results-composite} shows that students in both the groups struggled to answer some questions correctly. Figure~\ref{fig:all_questions}(c) shows one of the questions (Q9) that more students in the ML group answered correctly as compared to those in the Control group. 

\subsection{Evaluate Module}

The SA of the evaluate module contains 12 questions that assess students' ability to evaluate PCPs. 
The fifth chart from the top in Figure~\ref{fig:results-composite} shows the number of students who answered each question correctly. 
The majority of the students in the ML group performed well on most of the questions. Students in the ML group \textbf{perform much better} on Question 2 and 6 (64\% and 61\% accuracy respectively) than students in the Control group (47\% accuracy and 33\% accuracy). For Question 9 (Figure~\ref{fig:all_questions}(d)), very few students in both the teams answered the question correctly, but the students in the ML group performed better overall. 


\subsection{Create Module}

In the Create module, students demonstrated their ability to use PCPs for analysis tasks by \textit{creating} a PCP given a CSV file and then answering 5 questions about the chart they create. The bottommost chart in Figure~\ref{fig:results-composite} shows a range plot demonstrating that a large majority of the students in the ML group answered questions accurately, whereas \textit{students in the Control group made more mistakes than students in the ML group}. 



\begin{figure}[!t]
  \centering 
  \includegraphics[width=\columnwidth]{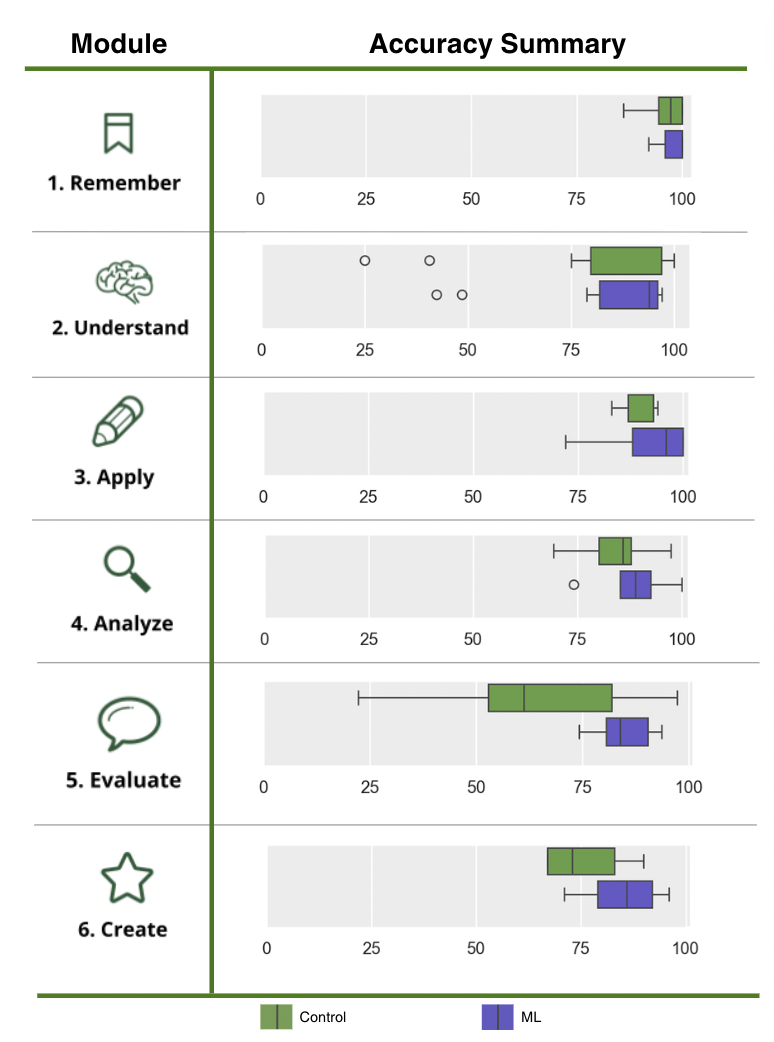}
\caption{A summary of the students' scores on the Summative Assessment (SA) for 
each of the six modules. Students in the ML group (purple) performed better in most modules, but the improvement is more pronounced in the advanced modules such as Analyze, Evaluate, and Create. }
\label{fig:modules-summary}
\end{figure}

\subsection{Results Summary} 

Figure~\ref{fig:modules-summary} shows a summary of the accuracy of students in every module. For the \textit{Remember} module, the accuracy of the students in the ML (purple) group was slightly higher with a median of 100, whereas the students in the Control (green) group had a slightly lower median score. Similarly, in the \textit{Understand} module, the distribution of accuracy of student scores is quite high in both the ML and Control modules. For the \textit{Apply} module, we see a higher median accuracy for the students in the ML group as compared to those in the Control group. For the \textit{Analyze} module, which is a more difficult module, we see that the students in the ML group performed better than the students in the Control group. 
Similarly, with the \textit{Evaluate} module, students in the ML group had a higher median score as compared to the students in the Control group. In the \textit{Create} module, students had higher accuracy with higher median scores for the students in the ML group. 

\begin{figure}[!t]
  \centering
  \includegraphics[width=\columnwidth]{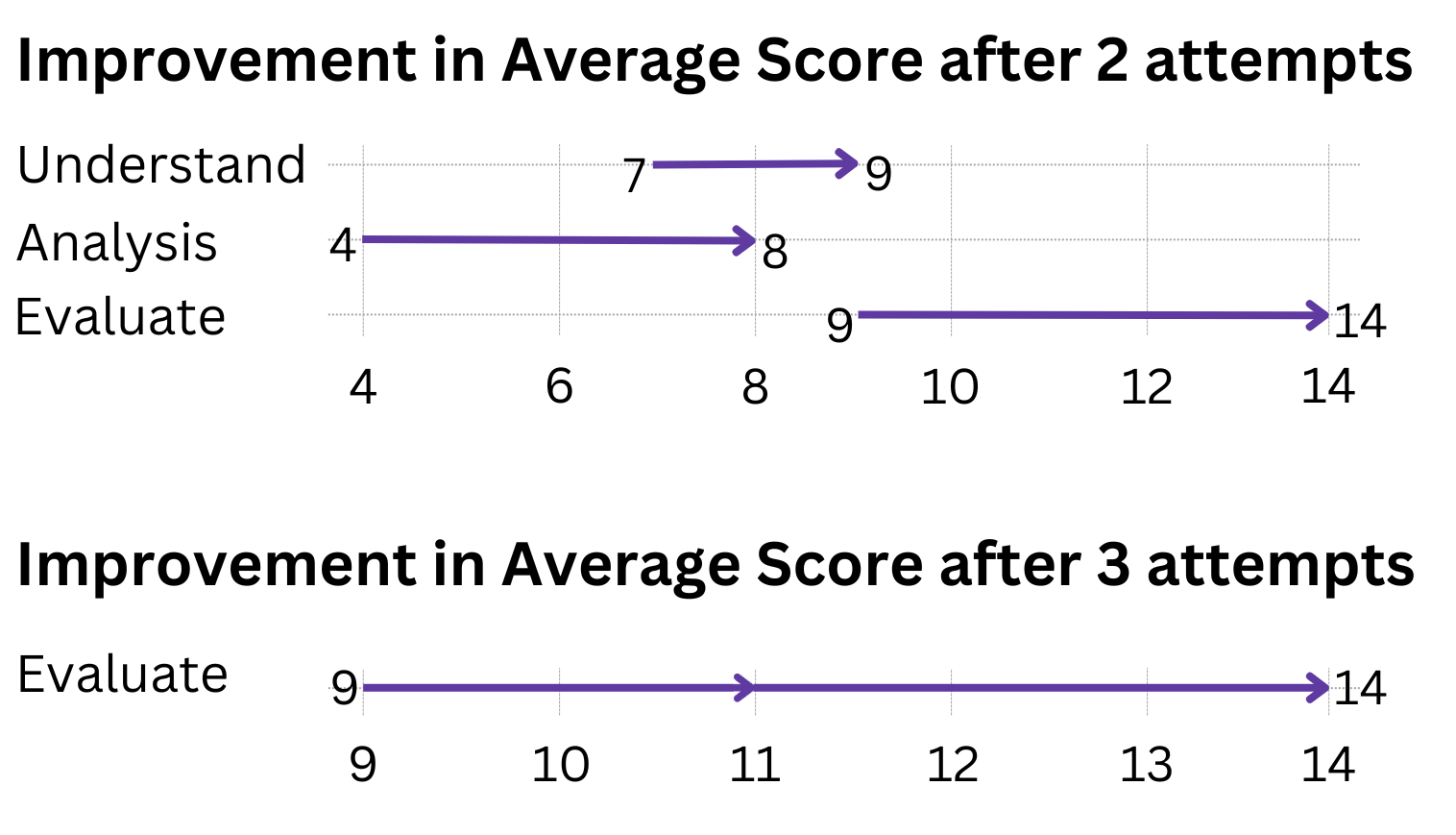}
  \caption{ML Improvements - Students scores are higher in the second iteration of the Understand, Analysis, and Evaluate FA. The
bottom chart shows the improvement in scores for the one student who repeated the Evaluate module three times.}
  \label{fig:mastery-learning-modules}
\end{figure}

\textbf{Impact of Retakes on Student Learning} - We examined the number of times students in the ML group had to retake the FA per module. While none of the students had to retake the FA for the Remember module, students had to retake the FA for the other modules. Figure~\ref{fig:mastery-learning-modules} shows the average improvement in scores for students in the ML group who retook the FA in each module. The top figure shows the average scores of students who took the FA once and received a low score, followed by a marked improvement in the second attempt. Average scores \textit{improved from 7 to 9 for the Understand module, from 4 to 9 for the Analyze module, and from 9 to 14 for the Evaluate module}. The bottom chart in Figure~\ref{fig:mastery-learning-modules} shows the improvement in the average score from 9 to 11 to 14 for the students who took attempted the Evaluate FA three times. There are no students who had to take the FA more than three times. Out of the 27 students in the Mastery group, \textit{only 1} student had to retake the FA for the Understand module, 5 students had to retake the FA for the Analyze module, and 13 students had to retake the FA for the Evaluate module. Only 2 students had to retake the Evaluate FA \textit{three times}. 

At the end of the intervention, we asked students to rate their confidence in being able to Recognize, Interpret, Create, and use PCPs in the future on data exploration tasks. Students in both the groups expressed their confidence using a 7-point Likert scale - 1 (low confidence) and 7 (high confidence). 
Students in both the groups conveyed high confidence with being able to \textit{recognize} and \textit{interpret} a PCP in the wild (7.0 for ML, 6.1 for control). The median student confidence to create a PCP is higher for the ML group  (7.0) as compared to the control group (5.0). Students in both groups conveyed high confidence about using it with a higher median (6.0) for the students in the ML group than those in the control group (5.0). 


Additionally, we measure student confidence in being able to recognize a PCP at three different stages in the intervention - (i) Before the intervention, (ii) after the Remember module, and (iii) at the end of all the six modules. 
The confidence improved from a \textbf{median of 3 to 6 or higher} for both the groups. 
This finding is encouraging, as it suggests that the Bloom's taxonomy-based modules enhanced students' ability to recognize PCPs. 


\subsection{Qualitative Feedback}
Students in both groups reported that they found the modules helpful in studying PCPs. 
When asked to improve the intervention, the students suggested using better tools
for the Apply task.
We allowed students to either hand-draw the PCP or use Google Slides to `draw' a PCP for the dataset that contained 6 rows of book information.
Some students in both groups suggested reducing the number of questions per module. One student expressed high interest in engaging with PCPs- ``\textit{It would be fun to upload our own data to try to create a parallel coordinate chart and play around with it.}''

When we asked students to \textit{compare a Scatterplot Matrix to a PCP representation} of the same data, we found that students could identify the advantages and disadvantages of both techniques. A student in the \textit{ML group} noted that PCPs make it'\textit {easy to follow the correlation of variables, but it gets very cluttered; hard to look at individual data points.}'' For the Scatterplot Matrix, they wrote~``\textit{it shows a good overview of each variable's relationship with itself and one another, but with many scatter plots to focus on; [it] may take up too much space.}'' A few more ML students highlighted PCPs' ability to show correlations across multiple variables. Still, they found that the ``\textit{clutter makes it hard to distinguish patterns.}'' 


The \textit{control group} students found that it was ``\textit{easier to visualize the correlation between two variables}'' when using a Scatterplot Matrix, but also commented that it was easier to see correlations across the dataset when using PCPs. One student summarized both techniques' drawbacks: - ``\textit{The scatterplot matrix shows poor visualizations when the number of variables increase (too many small multiples). The parallel coordinates chart gets visually cluttered when there are too many individual points.}'' These comments suggest that students understood PCPs well and could articulate the strengths and weaknesses of both visualization methods.

\section{Discussion}

Students in the ML group performed better in the higher-level thinking modules (Analyze, Evaluate, Create). The ML approach lead to an improvement in student learning based on the scores in subsequent attempts. Students in the ML group reported higher or equal confidence compared to the Control group in all areas related to PCP skills (recognize, interpret, create, and future use). Confidence in recognizing PCPs increased significantly for both groups throughout the intervention, demonstrating improved PCP literacy based on the Revised Bloom's taxonomy.

\subsection{PCP Literacy Test}

In addition to the details of the intervention, we have also open-sourced the Bloom's Taxonomy-inspired PCP Literacy Test (BTPL) that we developed. 
The \href{https://vis-graphics.github.io/PCP-Literacy-Test/}{BTPL} is an open educational resource that was designed to improve understanding and proficiency with PCPs. 
The test contains the videos developed, the questions included for the FA and the SA for each module of Bloom's taxonomy. 

The test can be used by an instructor teaching PCP in their classroom to introduce \textit{and} assess student learning after each module. We have provided links for students to watch the training video(s) and answer questions in the formative and SA on our website \protect\url{https://vis-graphics.github.io/PCP-Literacy-Test/}. 
The formative and summative assessments are applicable to the Remember, Understand, Analyze, and Evaluate modules. For the Apply and Create modules, we have included the original datasets used in our intervention for instructors and students to use, as well as the rubrics used to score the student created PCPs in the Apply and Create modules. 


Teaching PCP design through static images alone provides a limited training experience, as these plots often require dynamic exploration, and the interpretation of multidimensional data requires interaction (available in the Create module). The alignment with the Revised Bloom's taxonomy ensures that various cognitive levels are covered, further enhancing the deeper understanding of PCPs. 

This resource presents interactive demo videos and assessments to address these challenges and provides both basic and advanced skill acquisition in PCP comprehension and design interpretation. Furthermore, the FA provides direct feedback that has the potential to enable educators to identify learning gaps. For students, the testing process offers active engagement, iterative learning, and critical thinking opportunities. Its open-source nature makes it accessible, encourages independent practice, and fosters mastery of PCPs. This tool serves as a valuable pedagogical resource that helps overcome many limitations of classic teaching methods for the effective teaching and learning of PCPs.

\section{Summary}

We present the results of conducting a Bloom's taxonomy-based intervention and found that students in the group that had ML performed better than students in the Control group. The study suggests that the ML approach not only improved student performance in PCP tasks but also increased their confidence and understanding of PCPs. 
Additionally, we provide open PCP literacy modules for the 6 cognitive stages along with formative and summative assessments for reuse in the classroom. 


\bibliographystyle{IEEEtran}
\bibliography{references}

\begin{IEEEbiography}{Chandana Srinivas}{\,}is an undergraduate student in the Department of Computer Science at the University of San Francisco. Contact her at csrinivas2@usfca.edu.
\end{IEEEbiography}

\begin{IEEEbiography}{Elif E. Firat}{\,}is a Professor of Computer Science at Cukurova University. She received her Ph.D. from the University of Nottingham. Her research interest centers on information visualization and visualization literacy. Contact her at eefirat@cu.edu.tr.
\end{IEEEbiography}

\begin{IEEEbiography}{Robert S. Laramee}{\,}is a Professor of Computer Science and Data Visualization at the University of Nottingham. He was Associate Professor at Swansea University (Prifysgol Cymru Abertawe), Wales in the Department of Computer Science (Adran Gwyddor Cyfrifiadur). Contact him at Robert.Laramee@nottingham.ac.uk. 
\end{IEEEbiography}

\begin{IEEEbiography}{Alark P. Joshi}{\,}is an Associate Professor in the Department of Computer Science at the University of San Francisco. Dr. Joshi received his Ph.D. in Computer Science from the University of Maryland Baltimore County. His research has focused on data visualization and computer science education. Contact him at apjoshi@usfca.edu.
\end{IEEEbiography}



\end{document}